# Cryogenic Scintillation Properties of *n*-Type GaAs for the Direct Detection of MeV/c² Dark Matter

S. Derenzo[a], E. Bourret[b], S. Hanrahan[a], and G. Bizarri[a,c]

[a] Molecular Biophysics and Integrated Bioimaging Division, Lawrence Berkeley National Laboratory
[b] Materials Sciences Division, Lawrence Berkeley National Laboratory
[c] School of Aerospace, Transport and Manufacturing, Cranfield University, UK

## ABSTRACT

This paper is the first report of *n*-type GaAs as a cryogenic scintillation radiation detector for the detection of electron recoils from interacting dark matter (DM) particles in the poorly explored MeV/c² mass range. Seven GaAs samples from two commercial suppliers and with different silicon and boron concentrations were studied for their low temperature optical and scintillation properties. All samples are *n*-type even at low temperatures and exhibit emission between silicon donors and boron acceptors that peaks at 1.33 eV (930 nm). The lowest excitation band peaks at 1.44 eV (860 nm) and the overlap between the emission and excitation bands is small. The X-ray excited luminosities range from 7 to 43 photons/keV. Thermally stimulated luminescence measurements show that *n*-type GaAs does not accumulate metastable radiative states that could cause afterglow. Further development and use with cryogenic photodetectors promises a remarkable combination of large target size, ultra-low backgrounds, and a sensitivity to electron recoils of a few eV that would be produced by DM particles as light as a few MeV/c².

## I. INTRODUCTION

Despite overwhelming evidence that large amounts of cold DM are gravitationally associated with galaxies and galactic clusters, recent large-scale experiments designed to detect nuclear recoils from DM particles with masses above 1 GeV/c² have not yet seen a definitive signal [1-6]. This has motivated designs for experiments that search for DM particles in the MeV/c² range, using both electron and nuclear recoils. Calculations show that due to more favorable kinematics, light DM particles are expected to interact more efficiently with electrons than with nuclei [7-9]. For example, a DM particle with a mass of 10 MeV/c² and a typical velocity of $10^{-3}$ c has a kinetic energy of 5 eV, most of which can be transferred to a recoil electron and detected by low band gap detector materials. In contrast, the maximum energy that can be transferred by the same DM particle to a helium nucleus in an elastic collision is 0.05 eV, whose detection will require the use of ultra-low energy threshold processes such as the detection of helium atoms evaporated from super-fluid helium [10-12]. It is important to search for both electron and nuclear recoils since it is not known whether DM is leptophilic, hadrophilic, or both [13].

The calculations also show that energetic electron recoils are strongly suppressed and that the maximum detection rates will occur for electron recoil energies that are a small multiple of the band gap, even for DM masses as high as 1 GeV/c² [9]. As a result, the best detector materials for this application should have low bandgaps and low backgrounds. Scintillation detectors have an advantage over semiconductor ionization detectors (e.g. Ge) in that they do not require an electric field and afterglow from metastable radiative states is expected to be lower than dark currents [14]. This led to the choice of *n*-type GaAs for this work because of its low band gap (1.52 eV), donor-acceptor luminescence [15, 16], and the commercial production of large, high-quality crystals. An additional advantage is the apparent absence of afterglow, presented in Section III D of this paper.

This paper presents the first measurements of GaAs as a cryogenic scintillation radiation detector and is organized as follows: Section II describes the GaAs crystal samples and the measurement equipment used in this work. Section III presents experimental results and discussions of (1) optical emissions during optical and X-ray excitation, (2) afterglow, and (3) thermally stimulated luminescence. Section IV gives the conclusions of this work. The appendix uses previously published data to determine the minimum room temperature free carrier concentration in silicon-doped GaAs that maintains *n*-type conductivity at temperatures approaching 0 K (i.e. the Mott transition).

While this work focusses on the conversion of recoil electron energy into infrared photons, detection of those photons with high efficiency and low backgrounds at cryogenic temperatures is equally important. This technology is under active development using transition edge sensors [17] and microwave kinetic inductance detectors [18]. Another possibility is an optical absorber in superfluid helium. The heat from an absorbed 1.33 eV photon should produce about 1000 phonons and quasi-particles that will evaporate a corresponding number of helium atoms from the surface [12].

## II. SAMPLES AND MEASUREMENT EQUIPMENT

Section II.A describes the GaAs samples used and sections II.B to II.D describe the measurement equipment used. See Figure 1 for the experimental setup.

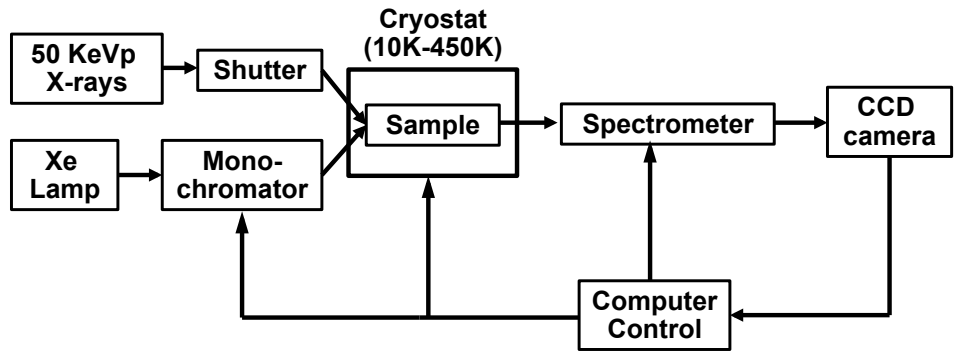

Figure 1. Experimental setup for the measurements reported in this paper.

## A. GaAs Samples

Table I lists the GaAs samples used in the measurements. They are in the form of wafers with thicknesses from 0.35 to 0.60 mm. All samples have low electrical resistances at 10 K, consistent with doping above the Mott transition concentration of 8 x $10^{15}$ free carriers cm$^{-3}$ (see Appendix). The exact *n*-type free carrier concentrations cannot be easily estimated from the silicon and boron dopant concentrations, since both can occupy gallium and arsenic sites.

## B. Cryostat

Samples were cooled with an ARS-2HW helium compressor and DE-202 two-stage cold finger (both from Advanced Research Systems, Macungie, PA). The temperature was regulated with a Model 336 controller (Lake Shore Cryotronics, Westerville, OH).

## C. Optically-excited Luminescence

Optical excitation was provided by a Model XS433 Xe broad-spectrum lamp coupled to a Model SP2150i double grating monochromator (both from Acton Research Corp., Acton, MA). A split optical fiber sent light from the monochromator to the sample and to a silicon reference photodiode. The photodiode was used to calibrate the excitation monochromator for lamp aging and for the wavelength-dependent combined output of the lamp, grating, and optical fiber. Fluorescence spectra were measured using a double-grating SpectraPro-2150i spectrometer with order-sorting filter wheel and a PIXIS:100B thermoelectrically cooled charge coupled detector (silicon CCD) (both from Princeton Instruments, Inc., Trenton, NJ). A white Teflon reflector was used to calibrate the emission spectrometer for variations in the combined response of the gratings, filters, and CCD as a function of wavelength. Samples were cooled using the cryostat described in section II.B.

Table I. GaAs samples used in this work. Si, B, and P concentrations determined by glow discharge mass spectrometry. Free carrier concentrations determined by Hall effect.

| Sample number | Supplier | Si ppm(wt) atoms/cm$^3$ | B ppm(wt) atoms/cm$^3$ | P ppm(wt) atoms/cm$^3$ | $n$-type free carriers/cm$^3$ |
|---|---|---|---|---|---|
| 13316 | AXT[a] | 8.9 $1.02 \times 10^{18}$ | 9.7 $2.87 \times 10^{18}$ | 0.008 $8 \times 10^{14}$ | $5.5 \times 10^{17}$ |
| 13352 | AXT[a] | 7.7 $8.8 \times 10^{17}$ | 11 $3.3 \times 10^{18}$ | 0.006 $6 \times 10^{14}$ | b |
| 13353 | AXT[a] | 7.2 $8.2 \times 10^{17}$ | 9.2 $2.73 \times 10^{18}$ | 0.003 $3 \times 10^{14}$ | b |
| 13354 | AXT[a] | 7.3 $8.3 \times 10^{17}$ | 12 $3.6 \times 10^{18}$ | 0.005 $5 \times 10^{14}$ | b |
| 13330 | Freiberger[c] | 8.3 $9.5 \times 10^{17}$ | 5.2 $1.54 \times 10^{18}$ | 2.8 $2.9 \times 10^{17}$ | $6.02 \times 10^{17}$ |
| 13332 | Freiberger[c] | 14.6 $1.67 \times 10^{18}$ | 6.8 $2.02 \times 10^{18}$ | 5.2 $5.4 \times 10^{17}$ | $1.44 \times 10^{18}$ |
| 13333 | Freiberger[c] | 20 $2.3 \times 10^{18}$ | 7.9 $2.34 \times 10^{18}$ | 4.4 $4.6 \times 10^{17}$ | $2.03 \times 10^{18}$ |

[a]AXT, Inc. (Fremont, California)
[b]Not measured but expected to be similar to sample 13316 since the concentrations of Si and B are similar
[c]Freiberger, Inc. (Freiberg, Germany)

## D. X-ray-excited Luminescence and Afterglow

The X-ray beam was produced by a Nonius FR591 water-cooled rotating copper-anode X-ray generator (50 kV, 60 mA) (Bruker AXS Inc., Madison, WI). The X-ray excited emission spectra, afterglow, and thermally stimulated emissions were measured using this system, the emission spectrometer described in section II.C, and the cryostat described in section II.B.

### III. RESULTS AND DISCUSSION

#### A. Optical excitation

Figure 2 shows the emission spectrum (horizontal axis) of sample #13316 at 10K as a function of excitation wavelength (vertical axis). A very weak emission band peaking at 850 nm (1.46 eV) and a much stronger emission band peaking at 930 nm (1.33 eV) are observed.

These emission bands are consistent with those reported in the literature. The emissions near 850 nm have been identified as transitions from shallow silicon donors to shallow acceptors [16, 19, 20], and the emissions near 930 nm as transitions from shallow silicon donors to boron acceptors on an arsenic site [16, 21]. Other emission bands centered near 1,050 nm and 1,300 nm (not presented in this work) have been associated with a gallium vacancy ($V_{Ga}$) bound to a donor [22] and a ($Si_{Ga}V_{Ga}Si_{Ga}$) complex [23], respectively. See [16] for a recent summary of the experimentally observed emissions, [24] for a discussion of the defects introduced by silicon and boron, and [20] for first-principles computations of their defect levels.

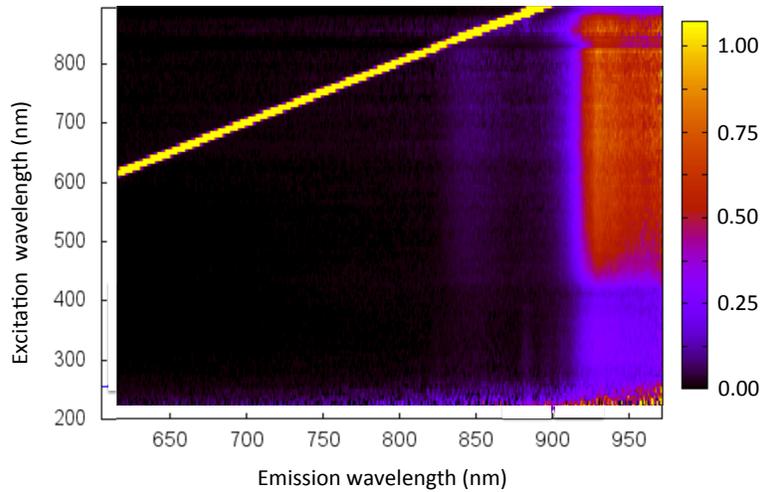

Figure 2. 2D map of emission as a function of excitation wavelength for sample #13316 at 10K.

Figure 3 shows the optical excitation and emission curves for sample #13316 at 10K. The excitation peak at 815 nm (1.52 eV) corresponds to the excitation of valence band electrons to the conduction band. The excitation peak at 860 nm (1.44 eV) corresponds to the excitation of boron acceptor electrons to the conduction band. The emission energy of 1.33 eV places the unoccupied boron acceptor level 0.19 eV above the valence band maximum, in good agreement with the previously measured value of 0.188 eV [21]. The Stokes shift between the lowest excitation energy peak and the emission peak is 0.11 eV. The overlap between the excitation and emission bands shows that photons with shorter wavelengths can be absorbed and re-emitted.

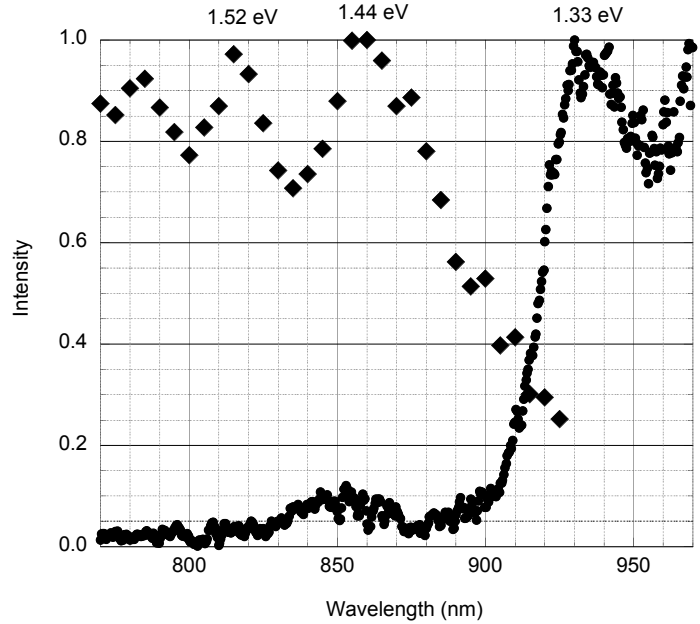

Figure 3. Sample #13316 excitation curve (diamonds) for 955 nm emission, and emission curve (circles) for 750 nm excitation at 10K.

The silicon donor ionization energy in GaAs is 2.3 meV [25]. Since boron is introduced during the growth of *n*-type GaAs crystals, at room temperature (k$T$ = 25 meV) some of the silicon donor electrons fill the boron acceptor sites and the remainder are distributed nearly equally between silicon atoms and the conduction band. The free carrier concentration for *n*-type GaAs is measured and reported under these conditions. On cooling to 0 K, free carriers below the Mott transition concentration bind to silicon atoms (i.e. are "frozen out") and the rest remain in the conduction band. Consequently, the concentration of free carriers at 0 K is equal to the free carrier concentration at room temperature minus the Mott transition concentration (see Appendix).

Figure 4(a) depicts the processes that occur when a 1.52 eV photon is absorbed: (1) A photon promotes a valence band electron into the conduction band, leaving a hole in the valence band, (2) a boron electron drops into the valence band hole, ionizing the boron, (3) the change in charge of the boron atom causes a lattice relaxation, (4) a donor electron combines with the ionized boron, producing a 1.33 eV photon. The radiative center could be an ionized boron acceptor and a single weakly bound donor electron (i.e. an exciton), or an ionized boron acceptor weakly bound to several equivalent donor electrons. Figure 4(b) depicts the processes that occur when a 1.44 eV photon is absorbed: (1) A photon promotes a boron electron into the conduction band followed by the previous processes (3) and (4). Not shown are mid-gap electron traps that are centers for non-radiative recombination [26]. As described in [24] boron can replace gallium as an isoelectronic substitution or it can replace arsenic as an acceptor.

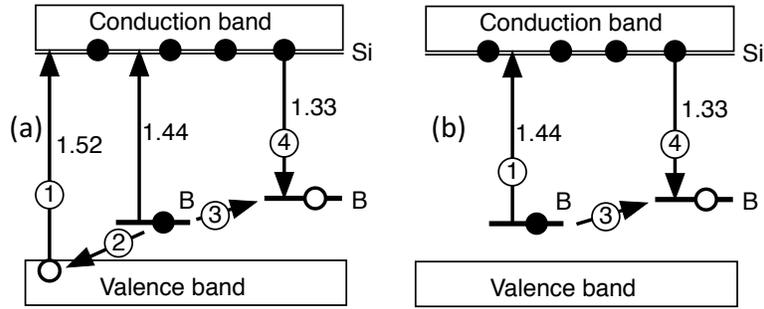

Figure 4. Simplified diagram of excitation and emission processes. Energy units in eV. See text for details.

## B. X-ray Excitation

Figure 5 shows the emission spectrum of sample #13316 when excited with 50 keVp X-rays at 10 K. The X-ray excited emission band at 930 nm has almost the same shape as the optically-excited emission band in Figure 3, indicating that the same radiative centers are excited. The observed luminosity was estimated to be about 43 photons/keV by calibrating against standard $Bi_4Ge_3O_{12}$ and $Lu_2SiO_5$:Ce scintillation crystals of equal size mounted in the cryostat in the same way and correcting for the wavelength-dependent sensitivity of the emission spectrometer (described in section II C). The $Bi_4Ge_3O_{12}$ and $Lu_2SiO_5$:Ce crystals were used at room temperature where their scintillation luminosities of 8 and 30 photons/keV, respectively, are well established [27, 28]. The value of 43 photons/keV for sample #13316 is an underestimate because (1) it does not include emissions above 970 nm and (2) the higher refractive index of GaAs results in more internal trapping than the standards. The index of refraction is 3.45 at 930 nm [29] and anti-reflection coatings can be used to reduce internal trapping of the scintillation light [30].

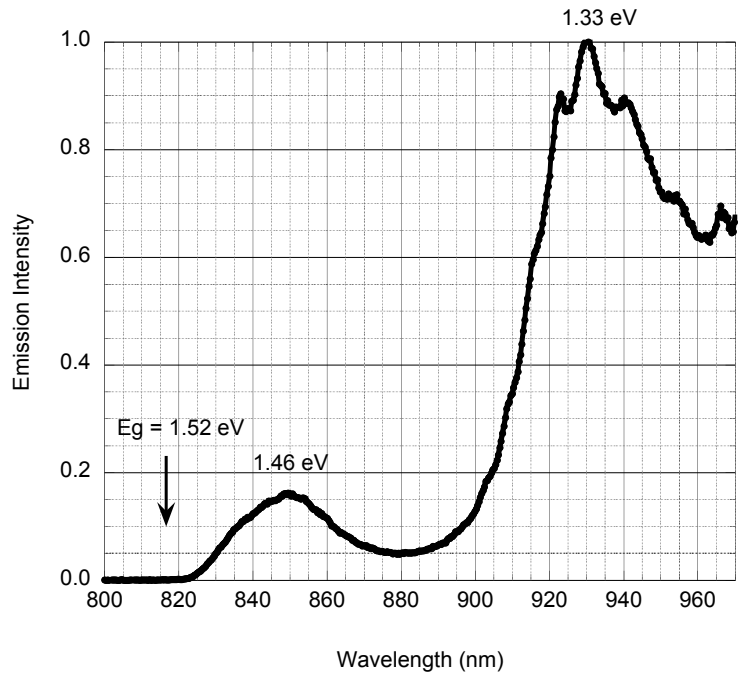

Figure 5. 50 keVp X-ray excited emission spectrum of sample #13316 at 10K

Figure 6 shows the thermal quenching of the 850 nm and 930 nm emission bands during 50 keVp X-ray excitation. The 850-nm emission is quenched above 20 K with an activation energy $E_a = 4.3$ meV. The 930 nm emission is quenched at 120 K with an activation energy $E_a = 12$ meV. We conclude that the latter value is the thermal barrier for the transfer of a hole from the boron acceptor to deeper traps [26].

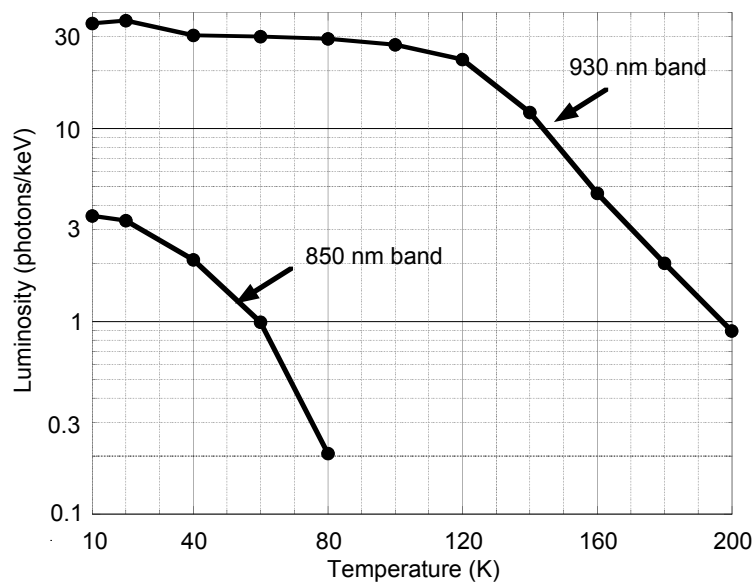

Figure 6. Luminosity of sample #13316 in the 930 nm and 850 nm emission bands during 50 keVp X-ray excitation as a function of temperature.

Table II lists the X-ray excited luminosity for the seven samples used in this work. While there is a weak correlation between doping levels and total luminosity of the AXT and the Freiberger samples taken separately, the primary difference is that the AXT samples are about five times more luminous than the Freiberger samples. The lower total luminosity of the Freiberger set could be due to a higher concentration of deeper impurity levels that are more efficient than boron in trapping holes [26], or could be due to the higher concentration of phosphorous. In the Freiberger set the emission shifts from 930 nm to 850 nm as the free carrier concentration is increased (e.g. sample #13330 vs. #13333). This indicates an increased involvement of shallow acceptor defects as the silicon concentration is increased. In a large crystal, an 850-nm photon would be absorbed and re-emitted into the 930-nm band (Figure 2).

The X-rays used in this work produced recoil electrons with a maximum energy of 50 keV. The range of a 50 keV electron is about 0.01 gm/cm$^2$, a value that is very weakly dependent on atomic number [31]. In GaAs (density 5.32 gm/cm$^3$) this corresponds to a range of about 0.02 mm, which is much less than the thickness of the samples used in this work.

Table II. Scintillation properties of seven samples at 10K.

| Sample number | Supplier | Luminosity (photons/keV) | Fraction @850 nm | Fraction @930 nm |
|---|---|---|---|---|
| 13316 | AXT | 43 | 0.08 | 0.92 |
| 13352 | AXT | 39 | 0.05 | 0.95 |
| 13353 | AXT | 33 | 0.08 | 0.92 |
| 13354 | AXT | 31.5 | 0.08 | 0.92 |
| 13330 | Freiberger | 7 | 0.33 | 0.67 |
| 13332 | Freiberger | 8.9 | 0.55 | 0.45 |
| 13333 | Freiberger | 9.5 | 0.85 | 0.15 |

## C. Afterglow

Figure 7 shows the scintillation intensity of sample #13330 at 10K in 2s time steps before, during, and after exposure to the 50 keVp X-ray beam. The beam shutter was manually opened from 300s to 900s. Since the closing of the X-ray shutter was not synchronized with the data acquisition, the point at 900 s could be an instrumental artifact.

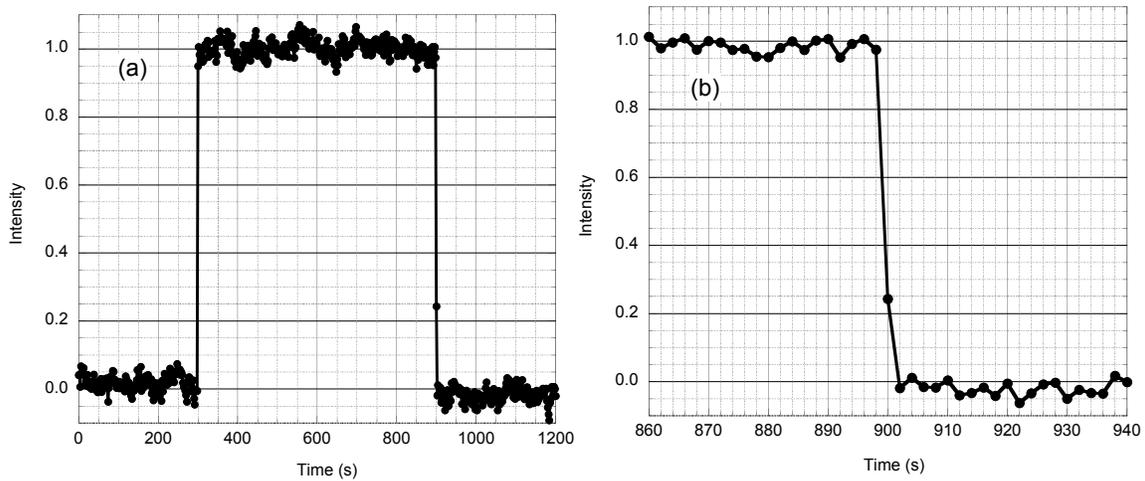

Figure 7. (a) Optical emission of sample 13316 at 10K before, during, and after exposure to a 50 keVp X-ray beam, measured in 2s time bins. (b) enlargement of emission around the time the beam was turned off.

## D. Thermally Stimulated Luminescence

Figure 8(a) shows the thermally stimulated luminescence from sample #13330 and crystals of Na(Tl), NaI, and CsI of similar size. The crystals were first excited with a 50 keVp X-ray beam for 30 minutes at 10K and then the emissions were recorded as the temperature was increased to 450 K over a period of 50 minutes. The emission peaks appear at temperatures characteristic of the trap depths of metastable excited states. If the temperature were held at 10 K after excitation, these same traps would slowly release to produce an afterglow background as the system returns to its ground state. A closer look at the GaAs emission curve Figure 8(b) shows that it is indistinguishable from the instrumental background. The apparent absence of thermoluminescence in *n*-type GaAs can be explained by the annihilation of metastable radiative states by the delocalized conduction band electrons. A search of the literature found only one paper describing thermally stimulated luminescence from a conductive semiconductor [32]. This paper described thermoluminescence from an n-type ZnSe crystal using UV (365 nm) excitation. This was later explained as the absorption of the UV exciting radiation (attenuation length $10^{-4}$ mm) by a layer of non-conductive surface defects [33].

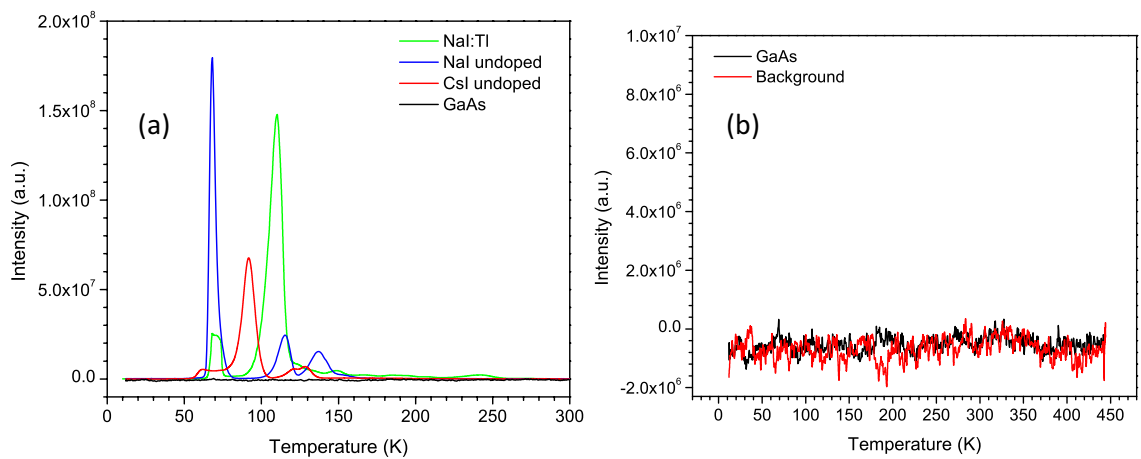

Figure 8. (a) Thermally stimulated luminescence of GaAs sample #13330 and crystals of NaI(Tl), NaI, and CsI as a function of temperature after a 30-minute exposure to a 50 keVp X-ray beam at 10K. All crystals were similar in size. The vertical scale is the same for all samples. (b) comparison between the GaAs curve and the instrumental background with vertical scale enlarged 20x

## IV. CONCLUSIONS

1) Electron excitations ≥1.44 eV can produce 1.33 eV photons.
2) The most luminous GaAs sample studied (#13316) has an observed scintillation luminosity of 43 photons/keV, below the theoretical limit of about 200 photons/keV.
3) Ionization from background radiation (e.g. muons and gamma rays) is not expected to cause afterglow, provided that the free carrier concentration is above the Mott transition, where even near 0 K conduction band electrons are not frozen out and can annihilate any metastable radiative states

These measurements show that *n*-type GaAs is a promising cryogenic scintillator for DM particle detection in the MeV/c² mass range in that it can be grown as large, high-quality crystals, has good scintillation luminosity, a threshold sensitivity at the 1.52 eV bandgap, and potentially no afterglow. However, more work is needed to optimize the doping concentrations, to reduce hole traps that compete with boron acceptors, and to develop anti-reflective coatings and cryogenic photodetectors. The *n*-type free carrier concentration must be high enough to efficiently combine with the boron acceptors and metastable radiative states but not so high as to introduce an unacceptable level of optical scattering and Auger quenching.

## Acknowledgements

We thank A. Canning, and R. Williams for discussions on scintillation mechanisms in semiconductors, M. Pyle for discussions on afterglow and optical absorption, T. Shalapska for assistance in data analysis, G. Martens for help in locating literature data, and Freiberger, Inc. for providing GaAs samples. This work was supported in part by the Office of Basic Energy Sciences of the U.S. Department of energy, in part by Advanced Crystal Technologies. Inc. of Knoxville, TN, and carried out using facilities provided by the U.S. Department of Homeland Security, Domestic Nuclear Detection Office at the Lawrence Berkeley National Laboratory under UC-DOE Contract No. DE-AC02-05CH11231.

## Appendix: The Mott transition in silicon-doped *n*-type GaAs

Figure 9 shows the electrical conductivity at 0 K of silicon-doped *n*-type GaAs as a function of the free carrier concentration measured at 300 K. The data are 0 K extrapolations taken from [25]. The conductivity is zero below 8 x $10^{15}$ cm$^3$ (the Mott transition) and rises linearly above that value. Below the Mott transition concentration, cooling from room temperature to 0 K causes the free carriers to become bound to individual silicon atoms at an energy level 2.3 meV below the conduction band minimum and they do not contribute to the electrical conductivity. Above the Mott transition concentration, cooling to 0 K results in a "metallic" state because mutual repulsion forces the additional electrons into the next higher available energy level, which is in the conduction band. At 0 K only the electrons forced into the conduction band contribute to the electrical conductivity and their concentration is equal to the room temperature carrier concentration minus the Mott transition concentration.

Exceeding the Mott transition concentration is important for scintillation at 0 K because it provides highly mobile conduction band electrons that can efficiently combine with ionization holes trapped anywhere in the gap. This serves to maximize the prompt radiative emission (the desired signal), and to annihilate metastable radiative states that could cause unwanted afterglow. However, an excessive free carrier concentration is not desirable because of increased optical scattering and Auger quenching.

It is interesting to note that both the donor ionization energy (2.3 meV) and Mott transition concentration (8 x $10^{15}$/cm$^3$) are uniquely low for GaAs(Si) [25]. In contrast, these values are 13 meV and 3.5 x $10^{17}$/cm$^3$ for Ge(As) [34]; 30 meV and 2 x $10^{18}$/cm$^3$ for CdS(Cl) [35]; and 45 meV and 3.7 x $10^{18}$/cm$^3$ for Si(P) [36].

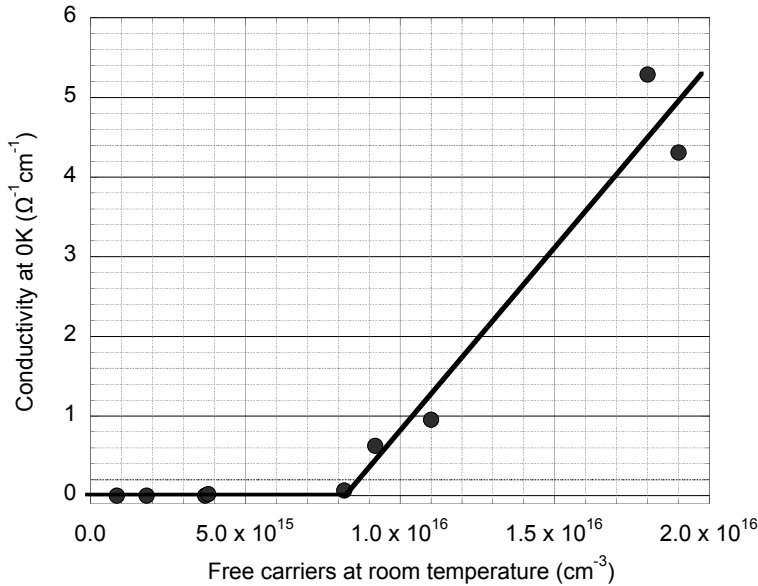

Figure 9. Electrical conductivity at 0 K as a function of the free carrier concentration at 300K, using data from [25]. When the room temperature free carrier concentration is below 8 x $10^{15}$ cm$^{-3}$ (the Mott transition), cooling to 0K causes the carriers to be bound to their silicon donor atoms (i.e. they are "frozen out") and the electrical conductivity is zero. Above the Mott transition, additional free carriers populate the conduction band and the electrical conductivity rises in proportion to their concentration.

**Disclaimer**